\documentclass[a4paper]{spie}  

\usepackage{amsmath,amsfonts,amssymb}
\usepackage{graphicx}
\usepackage[colorlinks=true, allcolors=blue]{hyperref}

\title{Crystal Eye: a wide sight to the universe looking for the electromagnetic counterpart of gravitational waves}

\author[a,b]{F.C.T. Barbato}
\author[b]{G. Barbarino}
\author[b]{A. Boiano}
\author[b]{R. de Asmundis}
\author[a,b]{F. Garufi}
\author[a,b]{F.Guarino}
\author[c]{R. Guida}
\author[c]{F. Renno}
\author[b]{A. Vanzanella}

\affil[a]{Department of Physics "E. Pancini", Via Cintia, Napoli, Italy}
\affil[b]{INFN - Section of Napoli, Via Cintia, Napoli, Italy}
\affil[c]{Department of Industrial Engineering, Piazzale Tecchio, Napoli, Italy}

\authorinfo{Further author information: (Send correspondence to F.C.T. Barbato)\\ E-mail: barbato@na.infn.it}

\begin{document} 
\maketitle

\begin{abstract}
With the observation of the gravitational wave event of August 17th 2017 and then with those of the extragalactic neutrino of September 22nd, the multi messenger astronomy era has definitely begun. With the opening of this new panorama, it is necessary to have a perfect coordination of the several observatories. Crystal Eye is an experiment aimed at the exploration of the electromagnetic counterpart of the gravitational wave events, that represent the missing observational link between short $\gamma$-ray bursts and gravitational waves from neutron star mergers. The experiment we propose is a wide field of view observatory. The Crystal Eye objectives will be: to alert the community about events containing soft X-ray and low energy $\gamma$-ray, to monitor long-term variabilities of X-ray sources, to stimulate multi-wavelength observations of variable objects, and to observe diffuse cosmic soft X-ray emissions.
\end{abstract}


\section{INTRODUCTION}
\label{sec:intro}  
A long-standing astrophysical paradigm is that collisions, or mergers, of two neutron stars create highly relativistic and collimated jets that power $\gamma$-ray bursts of short duration \cite{GW1, GW2, GW3}. The observational support for this model, however, was only indirect until August 2017, when the X-ray counterpart associated with the gravitational-wave event GW170817 \cite{GRB} was observed. The hitherto outstanding prediction that gravitational-wave events from such mergers should be associated with $\gamma$-ray bursts, and that a majority of these bursts should be seen off-axis (i.e. they should point away from Earth) was therefore verified.\\
In coincidence with this event, scientists performed observations at X-ray and, later, radio frequencies that result to be consistent with a short $\gamma$-ray burst viewed off-axis. The detection of X-ray emission at a location coincident with the kilonova transient provides the missing observational link between short $\gamma$-ray bursts and gravitational waves from neutron-star mergers, and gives independent confirmation of the collimated nature of the $\gamma$-ray-burst emission.\\
It results clear from the last observations that this is a new era for the observation of the Universe and we now need all the best and more elastic instruments in order to observe even when and where we don't expect anything.

\section{The scientific case: from Fermi-GBM to Crystal Eye}
\begin{figure}[h!]
		\centering
		\includegraphics[scale=0.8]{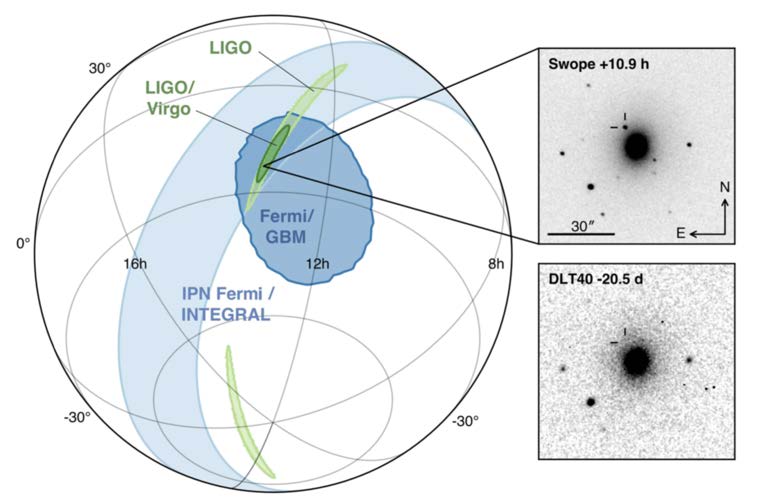}
		\caption{\emph{Localization of the gravitational-wave, gamma-ray, and optical 
		signals. it is synthetized the localization of the gravitational-wave, gamma-ray, and 
		optical signals. The left panel shows an orthographic projection of the 90\% credible 
		regions from LIGO (190 deg2; light green), the initial LIGO-Virgo localization (31 deg2; 
		dark green), IPN triangulation from the time delay between Fermi and INTEGRAL (light 
		blue), and Fermi-GBM (dark blue).}}\label{vista}
	\end{figure}

\begin{figure}[h!]
		\centering
		\includegraphics[scale=0.25]{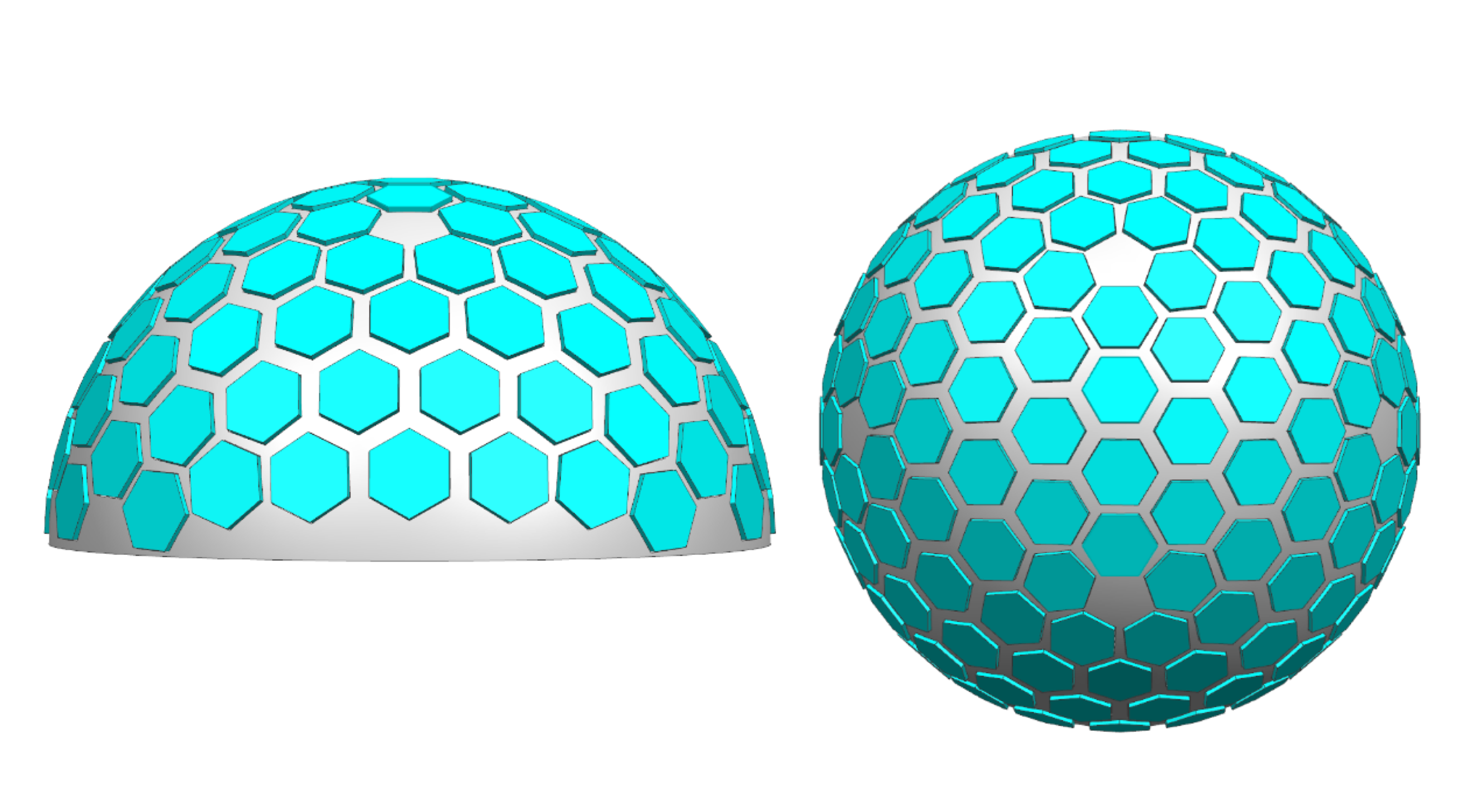}
		\tiny{\caption{\emph{Scheme of the Crystal Eye detector}}\label{CrystalEye}}
	\end{figure}

The main currently active experiments in orbit capable to observe the gamma and X-ray range are CHANDRA, XMM-NEWTON and FERMI. Despite their detection capability and their major observations, their concepts belong to two extreme sides of sky observatories.\\
On one side there are CHANDRA and XMM-NEWTON that have a high resolution camera but with a very small angle of view. This is their main limitation, as well as their main advantage. Indeed, if these characteristics  allow to point a source and to obtain a good imaging of that, on the other hand they limit the exploration of the Universe to a very small portion.\\
On the other side, we have all sky monitors as Fermi-GBM with a very wide field of view and a spatial resolution limited by its design.\\ 
In the FIG. \ref{vista}, there is the the first multimessanger observation made in 2017 \cite{2017}. It is evident from the figure, that the localization made by the Fermi $\gamma$-ray burst monitor instument (Fermi-GBM) on the FERMI satellite can be improved.
The GBM is the orbiting device currently in operation with the closest conceptual design with respect to Cristal Eye. Standing on the large difference of localization capability of the above mentioned class of experiments and to the vacancy in the observable energy range of those experiments, the Crystal Eye aims to be the joining link for the future observations, being an all sky pixelated monitor sensitive in the energy range going from tens of keV to few MeV.\\
The Crystal Eye (See FIG.\ref{CrystalEye}), will be a modern version of the Fermi-GBM detector, designed for a mission on the International Space Station (ISS) and thanks to its compactness can be usable as an optical module to be mounted on other satellites.
With this device, we aim to improve the localization capability of the GRBs by enhancing the spatial resolution of the monitor with a low-cost mission.\\
The requirements we set in order to make this project competitive, are the compactness (to be portable by an astronaut during a flight) and the wide field of view with small pixels (to improve the localization of the sources).\\
The Crystal Eye objectives will be: to alert the community about events containing X-rays and low energy $\gamma$-rays, to monitor long-term variabilities of X-ray sources, to stimulate multi-wavelength observations of variable objects and to observe diffuse cosmic X-ray emissions. The Crystal Eye will then be the most performant X and gamma-ray all sky monitor.

\section{The Crystal Eye method}
The device we are proposing is a semisphere made by a double layer of LYSO pixels covered by a veto dome. Around the Crystal Eye a low energy X-ray detector will be installed.
\subsection{The pixel}
\begin{figure}[h!]
		\centering
		\includegraphics[scale=0.35]{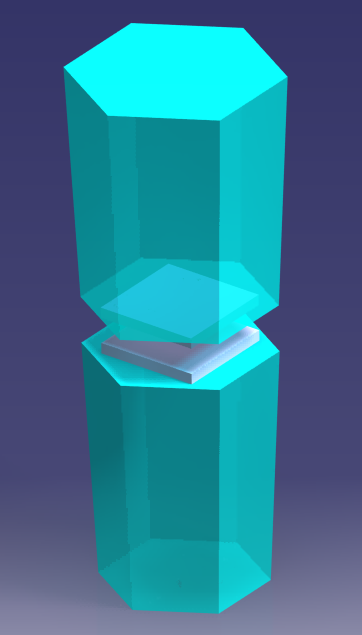}
		\tiny{\caption{\emph{Scheme of the Crystal Eye detector.}}\label{pixel}}
	\end{figure}

\begin{figure}[h!]
		\centering
		\includegraphics[scale=0.7]{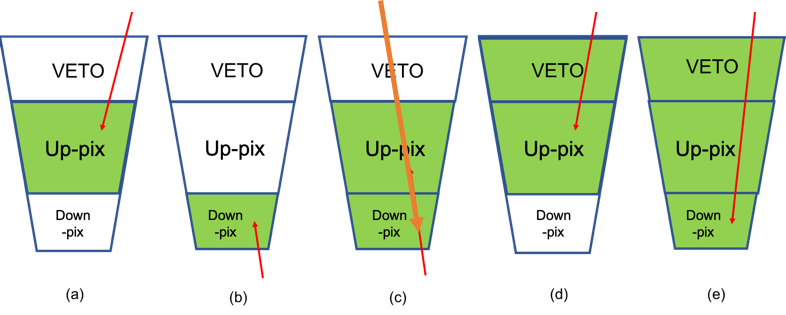}
		\tiny{\caption{\emph{Scheme of the Crystal Eye detector.}}\label{funzionamento}}
	\end{figure}

The device will have 110 pixels per layer. Each one is made by a hexagonal pyramid of LYSO read by a $12\times 12 mm^2$ SiPM array, see Fig. \ref{pixel}.
The pixel is 7cm high and the two scintillator layers will be respectively 3cm and 4cm high.
The hexagonal surface of the pixel has a diagonal of 3.2cm, almost four times less than the GBM pixel. The smaller size and the larger number of contiguous pixels will improve the spatial resolution and, as a consequence, the localization of the source.\\
The pixel is designed to observe the sky so to accept the down-going X and $gamma$-rays and to reject the up-going ones. Thanks to thickness of the pixel, if oriented along the direction of a down-going X or gamma ray with E $\leq$ 1 MeV, it will give a signal only in the upper pixel (FIG. \ref{funzionamento}a), while the up-going will give signal only in the lower one (FIG. \ref{funzionamento}b).\\
The external dome veto will allow the rejection of the charged particles crossing the detector (FIG \ref{funzionamento}d \ref{funzionamento}e). It will be made by a plastic scintillator with a dig where an optical fiber will guide the collected light to the SiPMs. The dome, used in anticoincidence with the crystals, will work as veto for the charged particles. The plastic scintillator is chosen in
order to be transparent to both X and gamma rays down-going.\\
Looking at the FIG. \ref{funzionamento}, the good events we are considering are those of the case (a).

\subsection{New materials for a compact detector}
\begin{figure}[h!]
		\centering
		\includegraphics[scale=0.25]{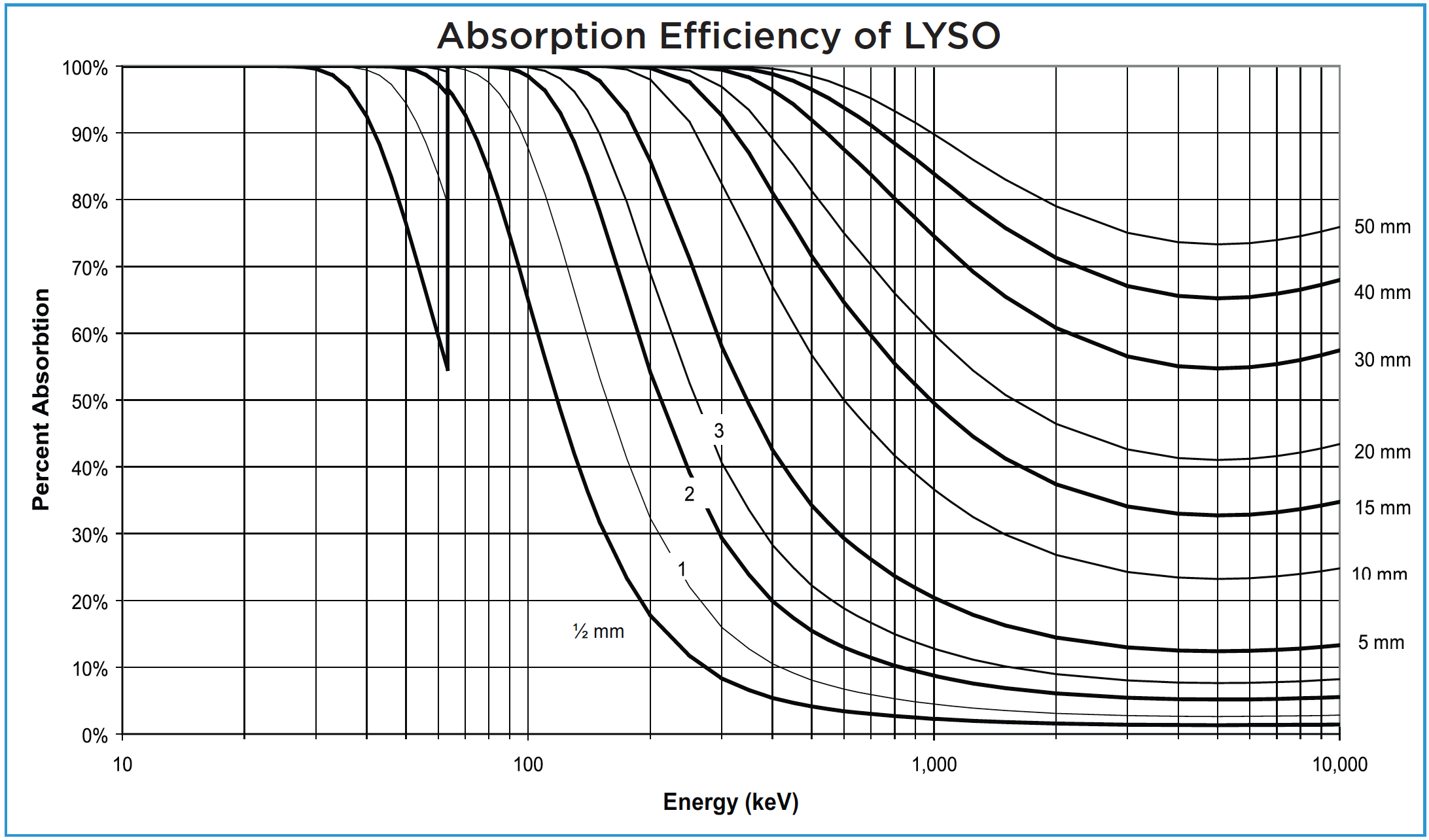}
		\tiny{\caption{\emph{Gamma and X-ray absorption efficiency
for various thicknesses of LYSO material.}}\label{LYSO}}
	\end{figure}

\begin{figure}[h!]
		\centering
		\includegraphics[scale=0.4]{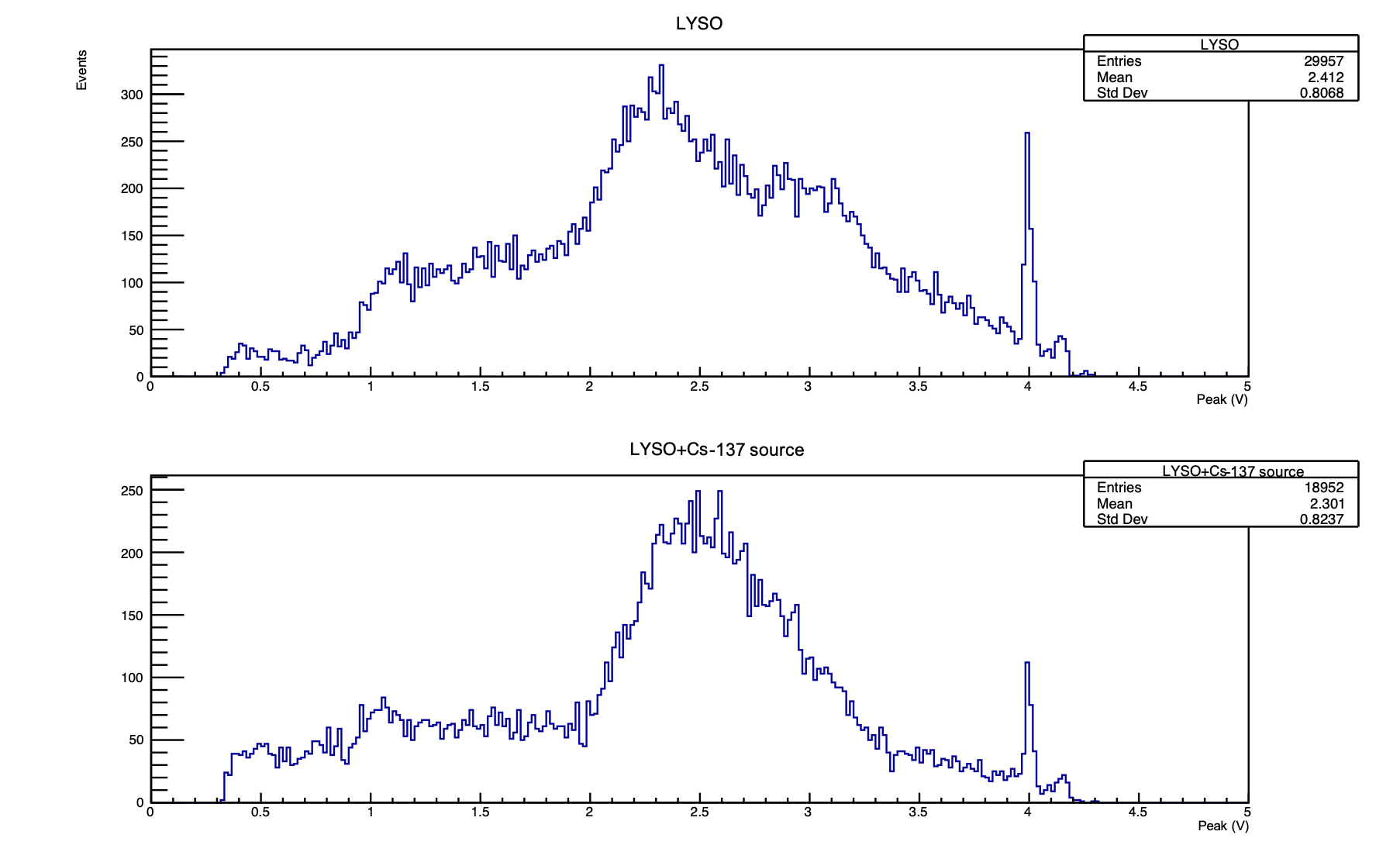}
		\tiny{\caption{\emph{TOP: LYSO radiactivity measured by MPPC13360-6075PE. BOTTOM: Cs-137 radioactive source signal in LYSO.}}\label{misureLYSO}}
	\end{figure}

\begin{figure}[h!]
		\centering
		\includegraphics[scale=0.55]{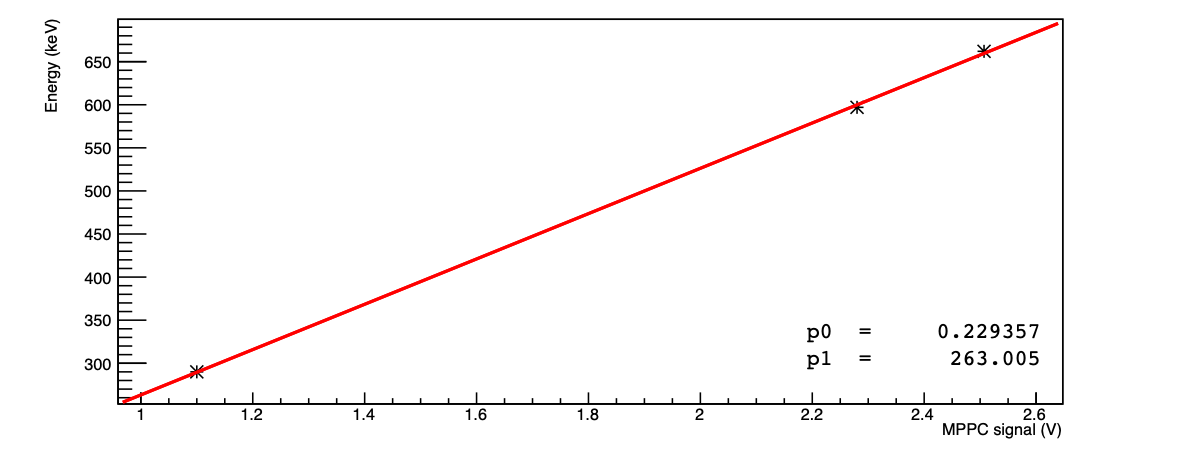}
		\tiny{\caption{\emph{Calibration in energy of the MPPC signal.}}\label{calib}}
	\end{figure}

The greatest innovation that Crystal Eye will introduce in the scientific panorama is from the technological point of view. The combined use of new sensors (SiPMs) and new materials for scintillators is the key to achieve very good detection features with a very compact and low cost device.\\
The hexagonal pyramids are made of LYSO. This new and very efficient material, see FIG. \ref{LYSO}, allows to detect $\gamma$-rays in the energy range of tens of MeV within few centimeters of scintillator.
LYSO is a Lutetium-based scintillator which contains a naturally occurring radioactive isotope 176Lu, a beta emitter. The decay results in a 3 $\gamma$-rays cascade of 307, 202 and 88 keV. This activity has been measured with a $6\times 6 mm^2$ MPPC S13360-6075PE and is represented by the highest peak reported in FIG. \ref{misureLYSO}TOP.
The activity has been estimated around 30counts/g.\\
An external Cs-137 source activity was measured (FIG. \ref{misureLYSO} BOTTOM) in order to effectively calibrate in energy the response of the MPPC, see FIG. \ref{calib}.
Moreover, we will exploit the internal radiactivity of the LYSO crystal as self-calibration of the MPPC array, when read in single mode.

\subsection{The electronics}\label{sec:elettronica}
\begin{figure}[h!]
		\centering
		\includegraphics[scale=0.45]{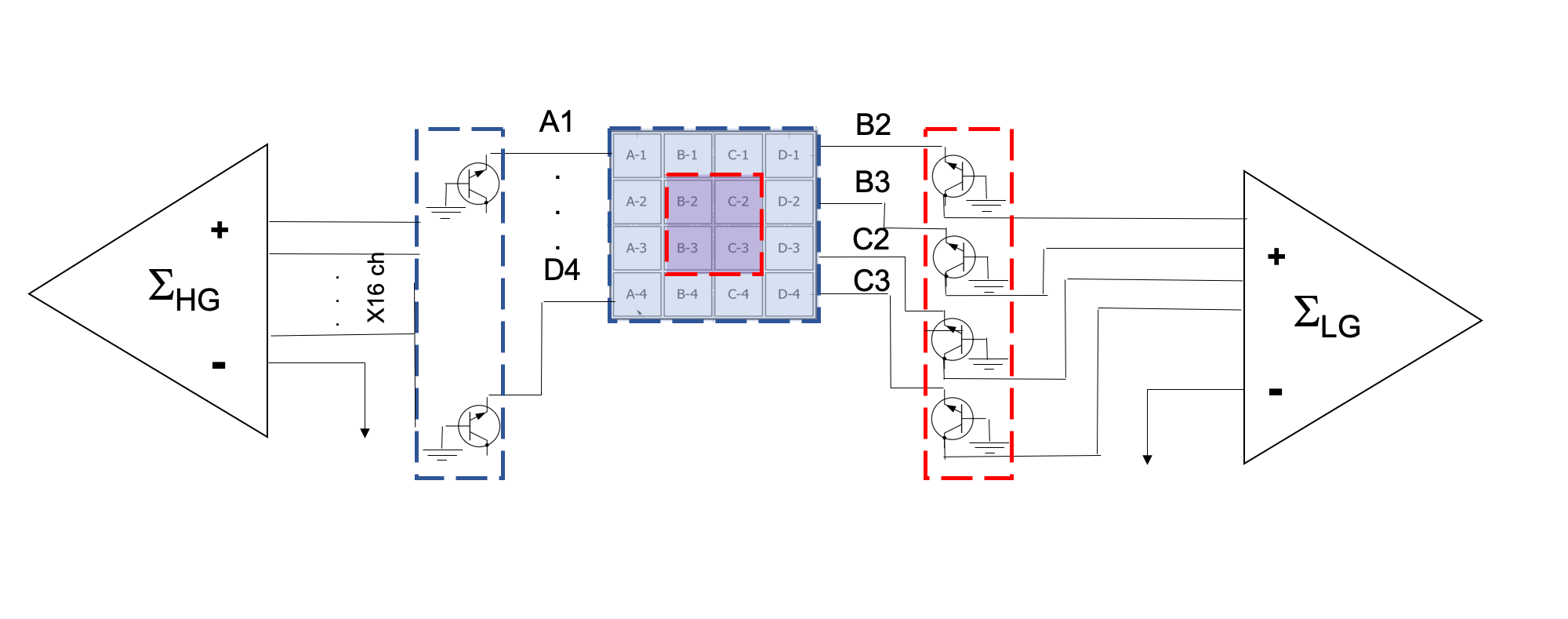}
		\tiny{\caption{\emph{Scheme of the frontend electronics.}}\label{arraysum}}
	\end{figure}

\begin{figure}[h!]
		\centering
		\includegraphics[scale=0.45]{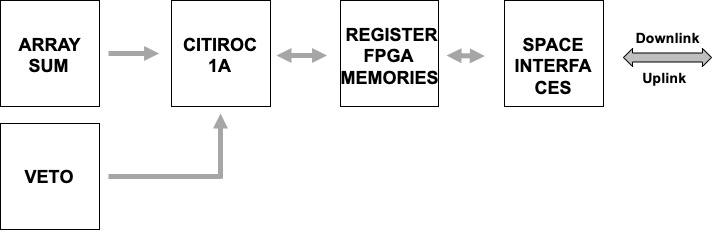}
		\tiny{\caption{\emph{Scheme of the electronics concept.}}\label{elettr}}
	\end{figure}

The pixel consists of two LYSO crystals, each read by an MPPC array S13361-3050. It is a 4$\times$4 matrix where the single element is an MPPC S13360-3050. The overall array is considered as a single readout channel. Since the amount of light on the MPPC array depends on the energy deposit in the crystal, we designed a front end electronics with two output lines: a low (LG) and a high gain (HG) output.\\
The LG output is obtained with the sum of the four central elements of the array and will be adopted for the high energy deposit. The HG output is obtained with the sum of all the elements of the array and will be adopted in case of low energy deposit, see FIG. \ref{arraysum}.
The array sum outputs will be the input channels of the CITIROC-A asic, developed by Omega, see FIG. \ref{elettr}.
Citiroc will manage the majority trigger and the anticoincidence of the veto. A topological trigger will be managed by the FPGA.\\
A GRB is expected when a majority of 5 up-pixels gives a signal above the selected threshold. We define this situation as the START signal of the GRB observation. By analizing the charge distribution among the pixel will be possible to localize the arrival direction of the GRB.

\section{The low energy detector: i-APS}
\begin{figure}[h!]
		\centering
		\includegraphics[scale=0.75]{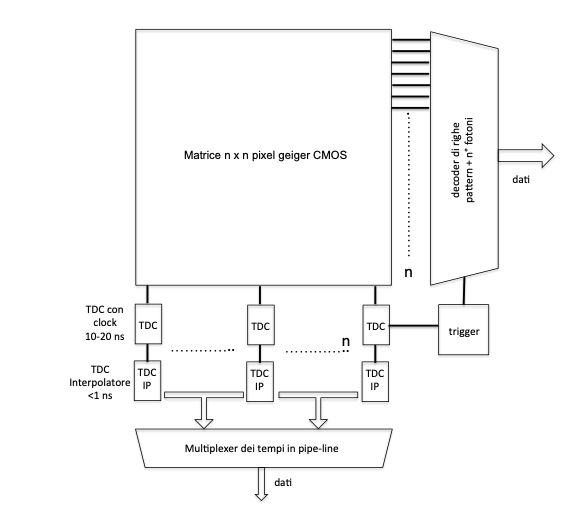}
		\tiny{\caption{\emph{Scheme of the i-APS concept.}}\label{i-APS}}
	\end{figure}

\begin{figure}[h!]
		\centering
		\includegraphics[scale=0.8]{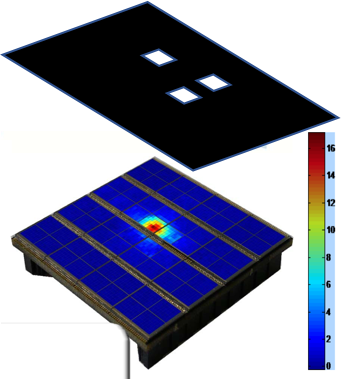}
		\tiny{\caption{\emph{Scheme of the coded mask concept.}}\label{maschera}}
	\end{figure}

The localization method described in \ref{sec:elettronica} fails when $E_{\gamma} \leq 200$ keV. In this situation, the charge will be always uniformly distributed among the pixels and it will be impossible to use this method for the localization of the burst. For this reason, we are developing a low energy detector: the imaging-Avalanche Pixel Sensor (i-APS). It is an imager made by an array of SPADs (Single Photon Avalanche Detectors), a technology in rapid evolution. The i-APS will be configured in order to detect photons in the energy region going from few keV up to hundreds of keV. The new i-APS structure aims to achieve a high spatial resolution and above all an excellent temporal response (100-200 ps). This is achieved thanks to a "parallel" acquisition architecture, differently from the CCDs. Each pixel is an inverse p-n junction, working in the Geiger region, associated with a single and independent output, see FIG. \ref{i-APS}. 
Thanks to this important difference, i-APS matrices will allow not only high speed image processing (100MHz), but also real time intelligence logic operations such as triggers and coincidences. The high internal gain entails a very high sensitivity, allowing the detection of the single photon.\\
The i-APS will be equipped with a coded mask, in this way for the low energy photons will be possible to localize the incoming direction thanks to the shape of the shadow given by the coded mask, see FIG. \ref{maschera}.

\section{Conclusions}
The Crystal Eye project started in March 2019 and aims at becoming the next future all sky monitor for X and gamma-rays.\\
It's design made by a pixelated semisphere orbiting around the Earth is realized to improve the localization of the sources of gravitational waves.\\
The goal will be achieved thanks to the use of new materials and sensors with respect to those currently in use in the ongoing orbiting experiments. Radiation hardness measurements arge ongoing. The pixels will be realized by using LYSO scintillators read by MPPC arrays. The high density of the material, together with the miniaturization of the sensor allow to realize a high granularity detector. \\
A flight prototype will be completed by 2021 in order to be launched for a 2 months flight test with Space Rider in the early 2022.

\section*{Acknowledgments}
The Crystal Eye project research is carried out in the frame of Programme STAR, financially supported by UniNA and Intesa San Paolo.\\
The i-APS project is financially supported by Italian Space Agency.

\section*{References}
\bibliography{report} 
\bibliographystyle{spiebib} 

\end{document}